\documentstyle[aps,prb,epsfig,manuscript]{revtex}
\begin{document}
\title{Exact multi-electronic electron-concentration dependent ground-states 
for disordered two-dimensional two-band systems in presence of disordered
hoppings and finite on-site random interactions.}
\author{Zsolt Gul\'acsi}
\address{Department of Theoretical Physics, University of Debrecen,
H-4010 Debrecen, Poroszlay ut 6/C, Hungary. }
\date{October, 2003}
\maketitle
\begin{abstract}
We report exact multielectronic ground-states dependent on electron 
concentration for quantum mechanical two-dimensional disordered two-band type 
many body models in the presence of disordered hoppings and disordered 
repulsive finite Hubbard interactions, in fixed lattice topology considered 
provided by Bravais lattices. The obtained ground-states loose their 
eigenfunction character for independent electron approximation, perturbatively
are not connected to the non-interacting but disordered case, and describe a 
localization-delocalization transition driven by the electron 
concentration, being highly degenerated and paramagnetic.
\end{abstract}

\section{Introduction}

In the real life, the crystallin state is the exception rather than the rule
\cite{bev1}, and as a consequence, the disorder exists everywhere, 
ranging from few impurities or interstitials in a periodic host, up to the
completely disordered glassy and amorphous structures, alloys and compounds. 
Given by this, the effects of the disorder are intensively analyzed 
\cite{bev2}, special attention being given in the last period to 
two-dimensional (2D) systems, where the observation of metallic behavior in 
2D high-mobility samples \cite{bev3} contradicting the
conventional non-interacting scaling theory \cite{bev4}, has underlined the 
special importance of electron-electron interactions in disordered systems, 
at least when its value is relatively high \cite{bev5}, or when the 
competition between disorder and interaction demands the consideration of
both \cite{tb1,tb2,tb3,tb4,tb5,tb6,tb6a}. 

Deep-rooted in the difficulty of describing the effects of the disorder in a 
non-approximated manner, on the theoretical side the interpretations are given
almost exclusively based on approximations. In the last decade however becomes
to be clear that this is not a fortunate situation, since not only the 
non-interacting scaling theory has been affected by new experimental results, 
but also other approximated schemes considered previously
indisputable (between them, all aspects of the Boltzmann description even for
the week-disorder limit in the treatement of the low temperature resistivity,
\cite{bev1}) have been forced to be re-analyzed. Based on these facts, 
suggestions to follow new roads have been made, underlining that the 
disordered materials cannot be understood by evading the real issue,
and forcing the disorder into a mold of procedures standard for ordered
systems \cite{bev1}. Furthermore, it has been stressed that the 
non-perturbative view on disorder could lead to significant advancement
in the understanding and description of such syatems \cite{tb5}. 
In the same time, several recent developments in the field require the 
non-approximated solution of the wave-equation for disordered and 
interacting systems as a key feature for a much deeper 
understanding of the emerging processes, and their interpretation, especially 
where it is expected, or experimentally is seen, that the electrons are 
maintaining their long-range phase coherence and retain their wave nature, as 
in the case of solid grains, short wires, dots,  mesoscopics, proximity to 
critical points, presence of long-range order, or of some kind of order in 
general \cite{bev2,bev2xx}, presence of quantum interferences \cite{bev2a}, 
etc.   

On this background, the first steps towards exact results for 
disordered systems have been made. On this line especially non-periodic models
of different type were analyzed. In these models, the non-periodicity is 
considered as introducing the effect of the disorder in the Schr\"odinger
equation, and is taken into account
in different ways, for example as non-analytic behavior in the potential
\cite{fib1}, incommensurate potential \cite{inc}, quasiperiodicity \cite{tb}, 
topological disorder connected to tesselation \cite{ra3}, local 
bond-orientational order \cite{qua}, etc, these possibilities presenting also 
interdependences between them. This way of describing the disorder cannot be 
considered as representing the level of simple toy models only, since besides 
the fact that real physical systems holding such properties are known 
\cite{bev2,fib1,inc,qua}, there are concrete cases where it is also known that
a such type of representation (for example through quasiperiodicity) give 
analogous behavior for the system as random or disordered potentials 
\cite{fib1,inc}.  

In $D=1$ the majority of studies leading to exact results were given for 
Fibanocci type of lattices \cite{fib1,fib2,fib3}, and the interested reader
will find more extended information on the 1D subject in review articles,
like Ref.\cite{bev2}. For $D > 1$, which is for 
interest in this paper, the first exact results have been obtained for 
quasicrystalline systems, where first, theoremes dealing with structure have 
been formulated, such as those involving inflation rules, or Conway's theorem 
\cite{ra5,ra6}. 
Later on, in few cases, even  exact eigenfunctions have been deduced for 
Penrose lattice \cite{ra5} in 2D \cite{ra7,ra8,ra9,ra9a}. This type of lattice
being a prototype of quasicrystalline systems \cite{ujx1}, clearly exceedes 
the level of a curiosity of pure mathematical character since it is related to
nearest-neighbor bond orientational order which is observed for example in 
simulation of supercooled liquids and metallic glasses \cite{qua}, attracting
clear interest \cite{ujx2}. In the 2D Penrose lattice, in a simplified view, 
flat and thin rhombuses cover the plane completely, forcing the resulting 
pattern to be non-periodic and introducing disorder in the system. For these  
systems, in 2D exact eigenstates were obtained by  Kohmoto and 
Sutherland \cite{ra7} for a strictly localized state including in the 
Hamiltonian on-site disordered potential as well depending on the number of
bonds entering in a given site, Sutherland \cite{ra8} obtaining a self-similar
state taking into account as well on-site potential which may have eight 
different values depending on the nature of the site, Arai et al. 
\cite{ra9} obtaining new strictly localized states in comparison with 
those described in \cite{ra7}, and Repetovitcz et al. \cite{ra9a} obtaining 
eigenstates even in the presence of plaquette-diagonal hoppings. 
The knowledge of these results has clarified
puzzles related to the influence of the disorder in several aspects, in an
extent which only exact results can provide. We note on this line 
clarifications of disputes originating from the interpretation of numerical 
results \cite{ra7}, evidences for the self-similarity of some ground-states
in disordered systems \cite{ra8}, evidences for singular features of 
ground-states in certain non-periodical systems \cite{ra9}, occurence of
degeneracy proportional to the system size in eigenstates \cite{ra9}, scaling
properties of the exact ground-states \cite{ra8}, relative stability of the 
confined states on boundary conditions \cite{ra9}, possible existence of 
allowed and forbidden sites in the eigenstates \cite{ra9}, etc. 

We underline that the above mentioned exact eigenstates are valid only for 
independent electrons, e.g. they were deduced from models build up on a 
tight-binding Hamiltonian in ${\bf r}$ space describing a single electron 
moving on an aperiodic graph \cite{tb}. Because of this 
reason, and in light of the facts previously presented, it would be extremely 
stimulating for the field to see in what extent the deduced exact properties 
at independent electron level, remain or not valid in exact terms for a 
really multielectronic and interacting system as well in the presence of 
disorder. In our knowledge, exact results of this type, up to this moment, 
are not known.  

In this paper, we report for the first time exact ground-states depending on 
the electron concentration for multielectronic and interacting 2D systems in 
the presence of disorder. The ground-states are paramagnetic, lose their 
eigenstate nature in the independent electron approximation, present 
properties known in the Penrose-lattice (for example strong degeneracy 
proportional to the system size), describe a localization-delocalization 
transition driven by the electron concentration, and in the localized case 
present clear evidence for long-range phase coherence. 

The results are reported for two-band type models. 
The presence of two bands is not diminishing the applicability of the results
since, from one side, real materials are of 
multiband type, and the theoretical description is given usually by projecting
the multiband structure in a few band picture \cite{vol}, which is stoped only
for its relative simplicity at the one-band extreme level, when this is 
possible. From the other side, the experimental one, several materials
treated traditionally in a two-band picture have been experimentally found 
containing disorder and presenting extremely interesting properties (as 
non-Fermi liquid behavior for example \cite{goo}) whose emergence is 
considered connected to the presence of the disoreder (see for example 
\cite{ra13,ra14} and cited literature therein). 
Concerning the presence of real random systems holding two type of electrons,
we mention the intense activity related to rare-earth and actinide compounds
which behave as random Kondo insulators \cite{rpam,ra12,ra13} holding two type
of electrons ($d$ and $f$), whose properties are described in a fixed lattice 
topology, but randomly distributed Hamiltonian parameters \cite{ra14}. 

The procedure we use originates
from developments leading to the first exact ground-states for the 
periodic Anderson type models obtained at finite value of the interaction
in 1D \cite{go1,go2,go3}, 2D \cite{g1,g2,g3}, and 3D \cite{g4}, which have 
been made here applicable in the disordered case as well. Our model 
is build up on a 2D graph in 
${\bf r}$ space, whose all vertices are of the same rank (four edges
are collected by every vertex), so the topology is fixed. Four neighboring
nearest-neighbor vertices form elementary plaquettes, and hopping (including
the non-local hybridization as well) is possible along the edges and diagonals
of elementary plaquettes. On each vertex local on-site potentials are acting, 
and on each vertex local on-site Hubbard type repulsion is present as 
interaction. 

The remaining part of the paper is build up as follows. Section II. presents
the Hamiltonian, Section III. describes an exact transformation of the 
Hamiltonian which allows the deduction of the presented results, Sect.IV 
analyzes the disorder present in the system and provides concrete examples
for the emergence of the model conditions necessary for the solutions to 
occur, Sect.V. presents the exact ground-states, Sect.VI describes 
ground-state expectation values, and finally, Sect.VII. concluding the paper 
closes the presentation.

\section{The Hamiltonian of the model.}

The fixed topology of the described system allows us to treat the   
problem in a 2D tight-binding Hamiltonian defined in ${\bf r}$ space on a 
2D Bravais lattice with disordered Hamiltonian parameters. For this system 
we consider an unit cell $I$ described by the primitive vectors $({\bf x},
{\bf y})$, and we take into account two type of electrons denoted by the 
particle index $p$ as $p=d,f$. In these conditions our starting Hamiltonian 
has the form $\hat H = \hat H_0 + \hat H_{int}$, where
\begin{eqnarray}
&&\hat H_0 = \sum_{p=d,f} \sum_{p'=d,f} \sum_{\sigma} \left[ 
\sum_{{\bf r} \ne 0} (t^{p,p'}_{{\bf i},{\bf i} +{\bf r},{\bf r},\sigma} 
\hat p^{\dagger}_{{\bf i},\sigma} \hat p'_{{\bf i}+{\bf r},\sigma} + H.c.) +
t^{p,p'}_{{\bf i},{\bf i},\sigma} 
\hat p^{\dagger}_{{\bf i},0,\sigma} \hat p'_{{\bf i},\sigma} \right],
\nonumber\\
&&\hat H_{int}= \sum_{p=d,f} \sum_{\bf i} U^p_{\bf i} \hat n^p_{{\bf i},
\uparrow} \hat n^p_{{\bf i},\downarrow} \: .
\label{E1}
\end{eqnarray}
In the one-particle part $\hat H_0$, the ,,length'' of the hopping
denoted by ${\bf r}$ with possible non-zero values ${\bf x},{\bf y},
{\bf y}+{\bf x},{\bf y}-{\bf x}$, is allowed to extend only to distances
contained in $I$, i.e. nearest-neighbors (${\bf x},{\bf y}$) and 
next-nearest-neighbors (${\bf y}+{\bf x}, {\bf y}-{\bf x}$) (see Fig.1).
Denoting by $N_{\Lambda}$ the number of lattice sites in the system,
te random nature of $\hat H$ is given by a) the $2 N_{\Lambda}$ independent,
non-correlated, random (repulsive) on-site Hubbard interactions $U^p_{\bf i}$, 
$p=d,f$ contained in $\hat H_{int}$, and b) $2 N_{\Lambda}$ new independent, 
non-correlated and random $\hat H_0$ parameters chosen (as will be clarified 
below) from the (site, direction and spin dependent) $t^{p,p'}_{{\bf i},
{\bf i}+{\bf r},{\bf r}, \sigma}$ amplitudes. We underline that the 
$t^{p,p'}$ coefficients contain hybridization ($p \ne p'$), and on-site 
potential (${\bf r}=0$) terms as well. 

We demonstrate below, that for certain local conditions imposed for 
$t^{p,p'}$ parameters which maintain the number of $4 N_{\Lambda}$ independent 
non-correlated random variables in the system, the exact multielectronic 
ground-state wave function of $\hat H$ in the interacting case can be 
explicitly given in an electron concentration dependent manner.

We mention that the spin-dependent nature of the $\hat H_0$ parameters is not 
essential for our deduction. The $\hat H_0$ parameters {\em can be} in
principle spin dependent as well, and we underline this aspect, in order to 
extend the applicability of the results also to Hamiltonians with non-diagonal
hoppings \cite{mang,and1,and2} too. Furthermore, concerning the type of the 
model we use, we mention that for $U^d_{\bf i}=0$, the $\hat H$ from 
Eq.(\ref{E1}) represents a {\em disordered} periodic Anderson model (or 
Anderson lattice), while for $U^d_{\bf i} \ne 0$, Eq.(\ref{E1}) describe a 
disordered two-band Hubbard model. Our results are applicable in both cases. 
For physical realization of such type of systems see for example 
Ref. \cite{rpam}.
 
We further consider that the mobility of the two type of electrons present
in the system ($d$ and $f$) is different, and the ratio in mobility is the
same on all lattice sites. As a consequence, from the point of view of hopping
amplitudes, starting from amplitudes written for $d$ electrons, we have
\begin{eqnarray}
t^{p,p'}_{{\bf i},{\bf i}+{\bf r},{\bf r},\sigma} = w^{\delta_{p,f} + \delta_{
p',f}}t^{d,d}_{{\bf i},{\bf i}+{\bf r},{\bf r},\sigma} \: ,
\label{E2}
\end{eqnarray}
where $w$ is a (site independent) measure of the mobility ratios between 
$f$ and $d$ electrons. We mention that hopping amplitudes between different
orbitals often satisfies such type of relations in real systems \cite{and2}.

Concerning again the $t^{p,p'}$ terms, being interested in the behavior of
particles given by the disordered hoppings and interactions,
we only consider situations for which the localization of particles in local 
trapping centers is avoided, i.e. we have
\begin{eqnarray}
t^{p,p}_{{\bf i},{\bf i},0,\sigma} > 0 \: .
\label{E3}
\end{eqnarray}
In the following Section we are presenting a transformation of $\hat H$ in a
form that allows to obtain exact ground-states in its spectrum.

\section{The transformation of the Hamiltonian.}

Let us introduce a numbering of the lattice sites by the integer number $l$
in the studied 2D lattice containing $N_{\Lambda} = L \times L$ lattice sites,
strarting from the down-left corner in the lowest row $(l=1$), going from left
to right up to the end of the firts row $(l=L)$, then going upward and
continuing with the second row again from left to right, and so on. In this 
manner, for example around an arbitrary lattice site ${\bf i}$, numbered by 
$l=i$, we find the site
numbering notations presented in Fig.2.a. The introduced notation allow us to
turn from a vectorial site notation to a scalar one, which simplify as well the
notation of the Hamiltonian parameters. For example the $t^{p,p'}_{{\bf i},
{\bf i}+{\bf r},{\bf r},\sigma}$ for ${\bf r}={\bf x}$, (${\bf r}={\bf y}$)
at site $l=i$ becomes $t^{p,p'}_{i,i+1,x,\sigma}$, $(t^{p,p'}_{i,
i+L,y,\sigma})$ respectively (see Fig.2.a.). Similarly, the next-nearest
neighbor components $({\bf x}+{\bf y},{\bf y}-{\bf x})$ become
$t^{p,p'}_{i,i+1+L,x+y,\sigma}$, $t^{p,p'}_{i+1,i+L,y-x,\sigma}$.

Let us further introduce a plaquette operator  $\hat A_{{\bf i},\sigma}$ 
defined for every arbitrary cell $I_{\bf i}$ taken at site ${\bf i}$ 
(see Fig.2.b). The cell $I_{\bf i}$ is denoted by its down-left corner 
${\bf i}$. The sites inside $I_{\bf i}$, are numbered in a cell independent
manner by the index $n=1,2,3,4$ starting from the site ${\bf i}$ and
counting anti-clockwise inside the unit cell $I_{\bf i}$ (see Fig.2.b).
In these conditions we obtain for $\hat A_{{\bf i},\sigma}$ the expression
\begin{eqnarray}
\hat A_{{\bf i},\sigma} = \sum_{n=1}^4 (a_{n,d} \hat d_{{\bf i}+{\bf r}_n,
\sigma} + a_{n,f} \hat f_{{\bf i}+{\bf r}_n,\sigma}) \: ,
\label{E4}
\end{eqnarray}
where $a_{n,p}$ are numerical coefficients, the same in all unit cells 
(detailed description of this procedure can be found in Ref.\cite{g2}. For
unit cell independent notation of the coefficients $a_{n,p}$ see 
Ref. \cite{g4}). Let further connect to every unit 
cell $I_{l=i}$, two random variables $\epsilon_{i,\uparrow}$, and
$\epsilon_{i,\downarrow}$.

Our results are based on the observation that if we define the plaquette 
operator parameters $a_{n,p}$ via the non-linear system of equations
\begin{eqnarray}
&&t^{d,d}_{i,i+1,x,\sigma} = a^{*}_{1,d} a_{2,d} \epsilon_{i,\sigma} + 
a^{*}_{4,d}a_{3,d} \epsilon_{i-L,\sigma} \: ,
\nonumber\\
&&t^{d,d}_{i,i+L,y,\sigma} = a^{*}_{1,d} a_{4,d} \epsilon_{i,\sigma} + 
a^{*}_{2,d} a_{3,d} \epsilon_{i-1,\sigma} \: ,
\nonumber\\
&&t^{d,d}_{i,i+1+L,x+y,\sigma} = a^{*}_{1,d} a_{3,d} 
\epsilon_{i,\sigma} \: , 
\nonumber\\
&&t^{d,d}_{i+1,i+L,y-x,\sigma} = a^{*}_{2,d} a_{4,d} 
\epsilon_{i,\sigma} \: , 
\nonumber\\
&&t^{d,d}_{i,i,0,\sigma} = |a_{1,d}|^2 \epsilon_{i,\sigma} + |a_{2,d}|^2 
\epsilon_{i-1,\sigma} + |a_{3,d}|^2 \epsilon_{i-1-L,\sigma} + |a_{4,d}|^2 
\epsilon_{i-L,\sigma} \: ,
\label{E5}
\end{eqnarray}  
and $a_{n,f}= w a_{n,d}$ holds, where the parameter $w$ (see Eq.(\ref{E2}))
is real but arbitrary, then, taking into account periodic boundary conditions,
the one-particle part $\hat H_0$ of the starting Hamiltonian from 
Eq.(\ref{E1}) becomes
\begin{eqnarray}
\hat H_0 =
\sum_{{\bf i},\sigma} \epsilon_{{\bf i},\sigma} \hat A^{+}_{{\bf i},\sigma} 
\hat A_{{\bf i},\sigma} \: .
\label{E6}
\end{eqnarray}
Compairing the last equality of Eq.(\ref{E5}) to Eq.(\ref{E3}), we obtain
the condition $\epsilon_{i,\sigma} > 0$, although $\epsilon_{i,\sigma}$ are
random variables. As a consequence, $\hat H$ in Eq.(\ref{E1}) becomes 
positive semidefinite
\begin{eqnarray}
\hat H = \sum_{{\bf i},\sigma} \epsilon_{{\bf i},\sigma} \hat A^{+}_{{\bf i},
\sigma} \hat A_{{\bf i},\sigma} + \sum_{p=d,f} \sum_{\bf i} U^p_{\bf i} 
\hat n^p_{{\bf i},\uparrow} \hat n^p_{{\bf i},\downarrow} \: ,
\label{E7}
\end{eqnarray}
and this property preserves the potential possibility to obtain the explicit
form of the ground-state in the interacting case.

\section{The disorder in the system.}

\subsection{The presence of randomness in the model}

Before going further, we should analyze the kind of randomness we have in the
system. We start with the observation that $\hat H$ in Eq.(\ref{E7}) which
will be analyzed further on, is clearly disordered since contains  
$4 N_{\Lambda}$ independent, non-correlated (non-negative) arbitrary random 
variables $\epsilon_{i,\sigma}$ and $U^p_i$. However, the randomness must be 
understood not only at the level of the transformed Hamiltonian Eq.(\ref{E7}),
but also at the level of the starting $\hat H$ presented in Eq.(\ref{E1}). 
Since the disorder in $\hat H_{int}$ is the same in Eq.(\ref{E1}) and 
Eq.(\ref{E7}), this question relates only the randomness in $\hat H_0$. 
In order to understand the source of the disorder in $\hat H_0$ from 
Eq.(\ref{E1}), we have two different alternatives.

One possibility for this, is to observe the linear relationship between the 
on-site energy levels $t^{d,d}_{i,i,0,\sigma}$ and $\epsilon_{i,\sigma}$ in 
the last row of Eq.(\ref{E5}). As a
consequence, we can consider that the initial disordered parameters of
the starting $\hat H_0$ in Eq.(\ref{E1}) are the
$\eta_{i,\sigma}=t^{d,d}_{i,i,0,\sigma}$ variables whose number is 
$2 N_{\Lambda}$, and the $\epsilon_{i,\sigma}$ new disordered parameters 
from Eq.(\ref{E7}) are obtained from these by a linear transformation
\begin{eqnarray}
\eta_{i,\sigma} = |a_{1,d}|^2 \epsilon_{i,\sigma} + |a_{2,d}|^2 
\epsilon_{i-1,\sigma} + |a_{3,d}|^2 \epsilon_{i-1-L,\sigma} + |a_{4,d}|^2 
\epsilon_{i-L,\sigma} \: ,
\label{E7a}
\end{eqnarray}  
which contains also $2 N_{\Lambda}$ equations. Since the $\eta_{i,\sigma}$ 
disordered parameters are independent, in this view one can consider that  
$\hat H$, besides the randomness in $\hat H_{int}$, possesses as well 
,,diagonal-disorder'' at the level of $\hat H_0$ in Eq.(\ref{E1}).

Alternatively, one can consider in Eq.(\ref{E5}) the unit-cell diagonal
hopping amplitudes ($t^{d,d}_{i,i+L,y-x,\sigma}, t^{d,d}_{i,i+1+L,x+y,\sigma}$)
directly proportional to $\epsilon_{i,\sigma}$  
as the source of the disorder in the one-particle part of the Hamiltonian, 
$\hat H_0$ in Eq.(\ref{E1}). In this case, $\hat H$ is considered to contain 
besides the randomness in $\hat H_{int}$, also ,,non-diagonal'' disorder at the
level $\hat H_0$.

In both cases, the remaining equalities in Eq.(\ref{E5}) must be considered as
local constraints necessary for the solutions to occur. Since the number of
$\hat H_0$ parameters in Eq.(1) is much higher than the number $2 N_{\Lambda}$
of random one-particle variables, these constraints do
not alter the random nature of the disordered variables ($\{U^p_i,\eta_{i,
\sigma} \}$, or $\{U^p_i,\epsilon_{i,\sigma} \}$). Rather, they lead to
I.) interdependences between $\hat H_0$ parameters not containing the 
disordered variables, and II.) connect other $\hat H_0$ parameters to 
$\eta_{i,\sigma}$ or $\epsilon_{i,\sigma}$ disordered variables. These 
constraints emerge in the process of the transformation of Eq.(\ref{E1}) into
Eq.(\ref{E7}), and we underline that our solutions are valid only in the case
when this transformation can be done (i.e. Eq.(\ref{E5}) holds). Both cases
mentioned above as non-diagonal and diagonal disorder in $\hat H_0$ will be 
analyzed in details below.

\subsection{Connections to the solutions obtained for Penrose tiling.}

Considering the disorder in $\hat H_0$ as non-diagonal,
the here presented solutions can be viewed as arising from
extension of the conditions used in the exact study of the Penrose tiling. 
In order to understand this statement, let us introduce the constants
$K_1=a^{*}_{1,d} a_{3,d}$, $K_2=a^{*}_{2,d} a_{4,d}$,
and observe, that since $\epsilon_{i,\sigma}$ are random, the diagonal 
(next-nearest neighbor) hopping matrix elements
in every unit cell $I_i$, namely $t^{d,+}_{i,\sigma} = t^{d,d}_{i,i+1+L,
x+y,\sigma}$ and  $t^{d,-}_{i,\sigma} = t^{d,d}_{i+1,i+L,y-x,\sigma}$,
excepting their ratio ($K_1/K_2$), remain random as well
\begin{eqnarray}
t^{d,+}_{i,\sigma} = K_1 \epsilon_{i,\sigma} \: , \quad
t^{d,-}_{i,\sigma} = K_2 \epsilon_{i,\sigma} \: .
\label{E9}
\end{eqnarray}
Considering for example the hopping amplitudes without
directional dependence, i.e. $t^d_{i,2,\sigma}=t^{d,\pm}_{i,\sigma}$, and 
taking for simplicity $K_1=K_2=1$, we obtain
\begin{eqnarray}
t^d_{i,2,\sigma} = \epsilon_{i,\sigma} \: ,
\label{E10}
\end{eqnarray}
which (excepting the fixed sign of $\epsilon_{i,\sigma} > 0$) means completely
random and independent unit-cell diagonal hoppings for all spins in all unit 
cells (see Fig.3.). 
As a consequence, based on Eq.(\ref{E9}) or its particular form from 
Eq.(\ref{E10}), we see that the randomness given by $\{U^p_i,\epsilon_{i,
\sigma} \}$ in Eq.(\ref{E7}), can be considered to originate from the 
randomness given by $\{ U^p_i, t^{d,\pm}_{i,\sigma} \}$ at the level of the 
starting $\hat H$ presented in Eq.(\ref{E1}).
In this case, once the hopping amplitudes along the diagonals
of every unit cell have been randomly chosen, the remaining
$t^{p,p'}$ parameters can be determined based on them. The 
study of Eq.(\ref{E5}) shows that fixing the $t^{d,\pm}_{i,\sigma}$
values, we have the liberty to choose independently two more constants 
relating the one-particle part of the Hamiltonian $\hat H_0$, namely 
$K_3$, and $K_4$, (sign$\bar K >0$, $\bar K=K_2K_3K_4$), based on which 
\begin{eqnarray}
&&t^{d,d}_{i,i+1,x,\sigma} = K_3 t^{d,d}_{i-L+1,i,y-x,\sigma} + \frac{1}{K_3}
t^{d,d}_{i,i+1+L,x+y,\sigma},
t^{d,d}_{i,i+L,y,\sigma} = K_4 t^{d,d}_{i,i+L-1,y-x,\sigma} + \frac{1}{K_4}
t^{d,d}_{i,i+1+L,x+y,\sigma} \: ,
\nonumber\\
&&t^{d,d}_{i,i,0,\sigma} = \frac{K_1}{\bar K} t^{d,d}_{i,i+L+1,x+y,
\sigma} + \frac{\bar K}{K_1} t^{d,d}_{i-L-1,i,x+y,\sigma} +
\frac{K_4}{K_3} t^{d,d}_{i,i+L-1,y-x,\sigma} + \frac{K_3}{K_4} 
t^{d,d}_{i-L+1,i,y-x,\sigma} \: ,
\label{E11}
\end{eqnarray}
and the numerical coefficients present in Eq.(\ref{E5}) in function of $K_m$,
($m=1,2,3,4$), arbitrary parameters become
$a^{*}_{1,d} a_{3,d} = K_1, \: a^{*}_{2,d} a_{4,d} = K_2, \:
a^{*}_{1,d} a_{2,d} = K_1/K_3, \: a^{*}_{1,d} a_{4,d} = K_1/K_4, \:
a^{*}_{2,d} a_{3,d} = K_2 K_4, \: a^{*}_{4,d} a_{3,d} = K_2 K_3, \:
|a_{1,d}|^2=K_1^2 /\bar K, \: |a_{2,d}|^2 = K_2 K_4 / K_3, \:
|a_{3,d}|^2=\bar K, \: |a_{4,d}|^2 = K_2 K_3 / K_4.$
In order to have real value for all $t^{p,p'}$ parameters, all $K_m$ must be 
real. To understand in details Eq.(\ref{E11}), let us introduce short
notations as well for nearest-neighbor and local amplitudes in the form
${t}^{d,x}_{i,\sigma}=t^{d,d}_{i,i+1,x,\sigma}$,
${t}^{d,y}_{i,\sigma}=t^{d,d}_{i,i+L,y,\sigma}$,
${t}^{d,0}_{i,\sigma}=t^{d,d}_{i,i,0,\sigma}$, which represent
the $t^{d,\nu}_{i,\sigma}$ amplitudes for $d$ electrons with spin $\sigma$ in 
unit cell $I_i$ for $\nu =|{\bf r}|$. Using these notations, Eq.(\ref{E11})
becomes
\begin{eqnarray}
&&{t}^{d,x}_{i,\sigma} = K_3 t^{d,-}_{i-L,\sigma} +K_3^{-1}t^{d,+}_{
i,\sigma} \: , \quad
{t}^{d,y}_{i,\sigma} = K_4 t^{d,-}_{i-1,\sigma} +K_4^{-1}t^{d,+}_{
i,\sigma} \: ,
\nonumber\\
&&{t}^{d,0}_{i,\sigma}=R_1 t^{d,+}_{i,\sigma} + R_1^{-1}t^{d,+}_{i-L-1,
\sigma} + R_2 t^{d,-}_{i-1,\sigma} + R_2^{-1} t^{d,-}_{i-L,\sigma} \: ,
\label{E13}
\end{eqnarray}
where $R_1=K_1/\bar K$ and $R_2=K_4/K_3$. As shown in Fig.4., the
$t^{d,\nu}_{i,\sigma}$ amplitudes presented in Eq.(\ref{E13}) are determined 
by the $t^{d,\pm}$ unit cell diagonal amplitudes that surround $t^{d,\nu}_{i,
\sigma}$. For example, as seen from Fig.4.a, the $t^{d,x}_{i,\sigma}$ nearest 
neighbor hopping amplitude (full line arrow) is given by the $t^{d,+}_{i,
\sigma}$ and $t^{d,-}_{i-L,\sigma}$ random unit cell diagonal amplitudes 
(dotted line arrows) which start from the same site $i$ and intercalate
$t^{d,x}_{i,\sigma}$. Similar situation is present for $t^{d,y}_{i,\sigma}$
in $y$ direction (see Fig.4.b), while $t^{d,0}_{i,\sigma}$ as seen in Fig.4.c.
is determined by the four $t^{d,\pm}$ ,,plaquette-diagonal'' amplitudes that 
start from the same site $i$.

Concerning Eq.(\ref{E13}), we mention that in the study of disordered systems,
constraints (correlations) between bond and site    
properties are often considered. The constraints a priori introduced can be in
some cases even of long range type, as taken for example in the case of 
isotropically correlated random potentials \cite{cor1}, correlated networks 
\cite{cor2}, etc., and even calculation techniques have been developed in 
order to deal with ,,constrainted'' disorder, for example in the form of 
correlated random numbers \cite{cor3}, or random 
matrices with symmetry properties or holding constraints \cite{cor2}.
In our case, local constraints exist which connect the plaquette diagonal bond
hoppings (considered as the true independent random variables of $\hat H_0$), 
to edge (nearest-neighbor) bond hoppings and local one-particle
potentials. Since the plaquette diagonal bond can be unambiguously connected
to the plaquette, in the described case, random plaquette properties (i.e. 
random bonds connected to plaquettes), through local constraints presented in 
Eq.(\ref{E13}), fix nearest-neighbor or local amplitudes.

Conrete physical situations where in disordered systems the random 
plaquette properties determine nearest-neighbor or local amplitudes are also
known in the literature. For example, in the case of topologically disordered 
system of Caer type \cite{ra1} using random mosaics, very similar to Voronoi 
tessellation generated from disordered arrangement of particles \cite{ra2}, 
random flips of plaquette-diagonal bonds performed with a given probability, 
determine the local nearest-neighbor hoppings, and introduce in this way 
the disorder in the system \cite{ra3}. 
Concerning disordered on-site one-particle terms generated by  
random bonds connected to plaquette properties, we mention for example the 
Penrose lattice \cite{ra5,ra6} case, where the on-site one-particle 
potentials have been introduced by the local coordination number \cite{ra7}.
In the mentioned case, practically the random on-site one-particle 
potential at site $i$ is determined by
the number of bonds entering in the site $i$. Our on-site potential 
$t^{p,p}_{i,i,0,\sigma}$ given in Eq.(\ref{E13}) and presented in Fig.4.c. 
has clear similarities with this choice, since reduces to a such
type of behavior in the case in which all $t^{d,\pm}_{i,\sigma}$ unit cell
diagonal hoppings are equal, and, for example, for all $m$, $K_m=1$. The 
diference between Eq.(\ref{E13}) and Ref.\cite{ra7} from the point of view of 
the random on-site potential is that in our case, the on-site potential is 
determined by the {\em value} of the random bonds entering in the site, while 
in \cite{ra7}, by the {\em number} of the bonds entering in the site. 
So contrary to Ref. \cite{ra7,ra8,ra9}, where the study has been 
concentrated on the effects of the lattice topology alone, in this paper we 
analyse the problem in a fixed topology, concentrating on random $t^{p,p'}$ 
values. We further mention, that in the Penrose lattice case, when also the 
plaquette diagonal hopping amplitudes are taken into account at the level of 
exact independent electron eigenstates \cite{ra9a}, solutions are found only 
when constraints are present between hopping amplitudes.

Let us consider a concrete physical example in support of Eq.(\ref{E13}) which
demontrates as well that solutions deduced in the context of Penrose lattice
\cite{ra9a} use quite similar conditions. For this, let us take a simple 
spin-independent case $t^{p,p'}_{i,j,r,\sigma} = t^{p,p'}_{i,j,r,-\sigma}=
t^{p,p'}_{i,j,r}$, and consider a situation for which, randomly positioned A 
or B atoms in the middle of the elementary plaquettes providing the random unit
cell diagonal hoppings $t^{d,\pm}_{i,\sigma}=t^{d,\pm}_i$ introduce the 
randomness in $\hat H_0$. In this 
situation, $t^{d,\pm}_i$ is either $t^{d,\pm}(A)$ or $t^{d,\pm}(B)$, depending
on the type of atom situated in the middle of an unit cell. For this 
example Eq.(\ref{E13}) express the fact that the hopping amplitude along a bond
(nearest-neighbor hoppings $t^{d,x}_i$ and $t^{d,y}_i$) depends on the randomly
situated atoms $A$ and $B$ placed in the neighborhood of the bond, and that
the on-site energy of a given site ($t^{d,0}_i$) depends on the randomly 
positioned $A$ and $B$ atoms in the neighborhood of the site, which are 
physically quite acceptable conditions. The linearity of these interdependences
can be physically motivated by the small influence of the atoms $A$ or $B$
not situated directly on the bond or on the site. 
For the presented example, $t^{d,\pm}(A)$ ($t^{d,\pm}(B)$) in the Penrose
lattice case would correspond to the 
notations $d_1,d_2$, ($d_3,d_4$) used in Ref.(\cite{ra9a}). Furthermore, in
Ref.(\cite{ra9a}) $t^{d,x}_i = t^{d,y}_i = 1$ is considered, and our 
$t^{d,0}_i$ is denoted by $\epsilon_i$. The conditions in which solutions are
obtained for the Penrose lattice case (see Eqs.(3.9),(3.11),(3.13) in Ref.(
\cite{ra9a})) are in fact linear relations of the type of our Eq.(\ref{E13}). 
The 
main difference between our model and that of Ref.(\cite{ra9a}) at the level
of $\hat H_0$ is that in our case, the plaquettes described by $t^{d,\pm}(A)$ 
and $t^{d,\pm}(B)$ can emerge completely random, while in Ref.(\cite{ra9a}),
the ,,plaquettes'' (,,rhombi'') described by $(d_1,d_2)$ and $(d_3,d_4)$ emerge
only with the randomness allowed by the Penrose tiling. Because of this reason,
our solutions extend the exact solution possibilities known in Penrose lattice
case to non-quasicrystalline disordered systems even in the presence of 
electron-electron interaction. 

\subsection{The disorder seen in the one-particle part of $\hat H$
as diagonal-disorder.}

Considering the source of the disorder in $\hat H_0$ diagonal, the random 
parameters of the model become $U^p_i$ and $\eta_{i,\sigma}$. In this case
Eq.(\ref{E5}) requires two supplementary conditions to be satisfied as follows.
I.) The next nearest-neighbor hoppings surounding a nearest-neighbor hopping 
(see Fig.4.a,b, and first two equalities from Eq.(\ref{E13})) must be related 
at the level of hopping amplitudes. These conditions are not specific for the 
presented disordered model, but are rather connected to the method itself. 
Indeed, such conditions we find solving non-disordered cases as well
(see Ref.(\cite{g1,g2,g3,g4})), and the obtained hopping amplitude
ratios are delimitating parameter space regions where the obtained solutions 
are valid. II.) The next-nearest-neighbor (unit cell diagonal) hoppings 
starting from a given site $i$ are all together linearly related to the 
on-site energy level (considered disordered here) at the site $i$ 
(see the last equality from Eq.(\ref{E13}), or alternatively Eq.(\ref{E7a}), 
and Fig.4.c.). This local constraint, in this form, is specific for the 
random case studied here.

Even if the conditions I. and II. presented above seems to be quite specific 
at first view, we show below that they ar compatible with the presence of the
diagonal disorder on physical grounds. For this to be visible, we analyze a 
simple pedagogical spin-independent hopping case which is $x-y$ symmetric as 
well, so $t^{d,x}_{i,\sigma} = t^{d,y}_{i,\sigma} = t^{d,\nu=1}_{i,\sigma} = 
t^{d,1}_i$, $t^{d,\pm}_{i,\sigma} = t^d_{i,2,\sigma}=t^{d,\nu=2}_{i,\sigma}=
t^{d,2}_i$, and $K_3 = K_4=K_0, R_2=1$ (see Eq.(\ref{E13})). Let further 
consider for the study that 
the random on-site potential $\eta_i$ is created by the randomly positioned 
$A_{\tau}$ atom at site $i$ of the lattice with lattice spacing $a$, where the 
index $\tau$ fixes the type of the atom. In this manner, if the atom 
$A_{\tau}$ will be placed at site $i$, it creates the on-site energy level 
$\eta_i=\eta_{\tau}$. After this step, we must model the expression of the 
distance dependent hopping amplitude $t_i(r)$ for the electron which starts 
the hopping from $i$. Taking into account a simple exponential
distance decrease, we may simply take $t_i(r) = C_i B_i(Av) e^{
-\alpha r}$, where the constant $\alpha$ describes the distance decrease ($r 
\ne 0$ is considered). The
amplitude $C_i B_i(Av)$ is build up from the component $C_i = C_{\tau}$ 
depending on the energy level at site $i$ (the atom $A_{\tau}$ present at 
site $i$), and the average effect of all surounding atoms felt at site $i$
denoted by $B_i(Av)$. Since only $t^{d,1}_i$ and $t^{d,2}_i$ hoppings are 
considered, we have for the $t^{d,1}_i$, ($t^{d,2}_i$) case the $r=a$, 
($r=a\sqrt{2}$) argument value in $t_i(r)$. 

After these considerations, the conditions I. and II. mentioned above look as 
follows. Condition I. links together 3 hopping amplitudes for hoppings which 
start from the same site $i$ (see Fig.4.a,b, and the equalities relating
$t^{d,1}_i$ from Eq.(\ref{E13})), providing the condition
$e^{\alpha a (\sqrt{2}-1)} = K_0 + (1/K_0)$.
As can be seen, condition I. determines in fact the strength of the hopping
(parameter $\alpha$) through the constant $K_0$, and is not specific for the
random case, as mentioned above. Rather, it fixes the $t^{d,1}/t^{d,2}$ ratio
introducing limits for the valability of the solutions in the $T=0$ phase 
diagram of the starting Hamiltonian. As a consequence, we can further analyze 
condition II. considering $\alpha$ known parameter.    

Condition II. (Fig.4.c, and the equality relating $\eta_i=t^{d,0}_{i,\sigma}$
from Eq.(\ref{E13})) links together four next-nearest-neighbor hoppings which
again start from the same site. As a consequence, taking into account that the
atom $A_{\tau}$ is placed on the lattice site $i$, we find
$\eta_{\tau} = C_{\tau} B_i(Av)(2+R_1+(1/R_1)) e^{- \sqrt{2} \alpha a}$. 
The remaining $B_i(Av)$ coefficients must be deduced at each site from the
condition $t^{*}_{i,j}=t_{j,i}$. As can be seen, condition II., through the
parameter $R_1$ relates the disordered on-site energy level values to the 
hopping amplitude components $C_i$. 
  
As presented above, the conditions necessary to be fulfilled for the solutions
to emerge are present in disordered systems, being compatible to a truly 
acceptable physical background. Taking into account more complicated 
parametrizations for $t_i(r)$, the equation of
$\eta_{i,\sigma}=t^{d,0}_{i,\sigma}$ in Eq.(\ref{E13}) (since has in its right
side the same $r$ value), reduces to an equation for the amplitudes of the $r$ 
function in $t_i(r)$, while the remaining equalities in Eq.(\ref{E13}) 
determine the $t_i(a)/t_i(a \sqrt{2})$ ratios.

\section{The ground-state wave-function.}

Starting from the positive semidefinite structure of $\hat H$ in Eq.(\ref{E7}),
the ground-state wave function $|\Psi_g\rangle$ is obtained for
$\hat H |\Psi_g\rangle = 0$. Now, let us concentrate first to the $\hat H_0$
component of $\hat H$ presented in Eq.(\ref{E6}). Taking into account Eq.(
\ref{E4}), and as shown under Eq.(\ref{E5}), $a_{n,f}=w a_{n,d}$, where $w$ is
arbitrary but real, we realize that
\begin{eqnarray}
\hat A_{{\bf i},\sigma} = \sum_{n=1}^{4} a_{n,d}(\hat d_{{\bf i}+{\bf r}_n,
\sigma} + w \hat f_{{\bf i}+{\bf r}_n,\sigma}) \: ,
\label{E14}
\end{eqnarray}
so in the right side of $\hat H_0$ in Eq.(\ref{E6}) only operators of the
form $\hat O_{j,\sigma} =(\hat d_{j,\sigma} + w \hat f_{j,\sigma})$ are 
present. If now we define
\begin{eqnarray}
\hat {\bar O}^{\dagger}_{j,\sigma} = \hat d^{\dagger}_{j,\sigma} -
\frac{1}{w} \hat f^{\dagger}_{j,\sigma} \: ,
\label{E15}
\end{eqnarray}
which satisfies $\hat O_{j,\sigma} \hat {\bar O}^{\dagger}_{j',\sigma'} =
-  \hat {\bar O}^{\dagger}_{j',\sigma'}\hat O_{j,\sigma}$, then taking
$|\Psi\rangle = \prod_j [\hat {\bar O}^{\dagger}_{j,\sigma} + v_j
\hat {\bar O}^{\dagger}_{j,-\sigma}] |0\rangle$, where $\prod_j$ is taken over 
different (although arbitrary) lattice sites, $v_i$ are arbitrary coefficients,
and $|0\rangle$ is the bare vacuum with no fermions present, we obtain
\begin{eqnarray}
\sum_{{\bf i},\sigma} \epsilon_{{\bf i},\sigma} \hat A^{+}_{{\bf i},\sigma} 
\hat A_{{\bf i},\sigma} \: |\Psi \rangle = 0 \: .
\label{E16}
\end{eqnarray}
Since $|\Psi\rangle$ introduce fermions ($d$ or $f$) with arbitrary spin,
strictly on different sites, double occupancy is avoided, and
\begin{eqnarray}
\sum_{p=d,f} \sum_{\bf i} U^p_{\bf i} \hat n^p_{{\bf i},\uparrow} 
\hat n^p_{{\bf i},\downarrow} \: |\Psi\rangle = 0 \: ,
\label{E17}
\end{eqnarray}  
holds as well. Since the minimum possible eigenvalue of $\hat H$ in 
Eq.(\ref{E7}) is zero, the ground-state for arbitrary 
$N \leq N_{\Lambda}$, where $N$ represents the number of electrons within the 
system, becomes
\begin{eqnarray}
|\Psi_g \rangle = \prod_j^{N} [\hat {\bar O}^{\dagger}_{j,\sigma} + v_j
\hat {\bar O}^{\dagger}_{j,-\sigma}] |0\rangle \: .  
\label{E18}
\end{eqnarray}
In Eq.(\ref{E18}), the $\prod_j^{N}$ product must be taken over $j$ sites 
which can be arbitrary chosen, and different $j$ values must be related to
strictly different lattice sites. The ground-state wave function
of $\hat H$ given in Eq.(\ref{E7}) for  $N \leq N_{\Lambda}$ (i.e. at and 
below quarter filling) can be always written in the form of Eq. (\ref{E18}).
As a consequence, for $N=N_{\Lambda}$ the ground-state becomes
\begin{eqnarray}
|\Psi_g (N=N_{\Lambda})\rangle = \prod_{j=1}^{N_{\Lambda}} [
\hat {\bar O}^{\dagger}_{j,\sigma} + v_j
\hat {\bar O}^{\dagger}_{j,-\sigma}] |0\rangle \: .  
\label{E19}
\end{eqnarray}
For $N < N_{\Lambda}$, since the $j$ sites in Eq.(\ref{E18}) can be arbitrary 
chosen, the complete ground-state become
\begin{eqnarray}
|\Psi_g (N < N_{\Lambda})\rangle = \sum_{R_N} \{ \alpha_{R_N} 
\prod_{j \in R_N} [
\hat {\bar O}^{\dagger}_{j,\sigma} + v_j \hat {\bar O}^{\dagger}_{j,-\sigma}] 
\} |0\rangle \: ,  
\label{E20}
\end{eqnarray}
where the sum $\sum_{R_N}$ is made over all different $R_N$ domains containing
$N < N_{\Lambda}$ lattice sites from the system, and $\alpha_{R_N}$ are
numerical coefficients.
Furthermore, it is important to underline that Eqs.(\ref{E19},\ref{E20}) 
represent the ground-state only in the interacting case
(at least one of on-site two-particle interactions $U^p_i$ must be non-zero 
at all sites $i$, since otherwise,
because of the presence of the double occupancy, $|\Psi_g\rangle$ in 
Eq.(\ref{E18}) is no more an eigenstate of the Hamiltonian).

In my knowledge, Eqs.(\ref{E19},\ref{E20}) contain the first exact 
multielectronic ground-state wave-functions obtained in 2D for a disordered 
system in the interacting case.  As explained above, 
Eqs.(\ref{E19},\ref{E20}) are {\em no more eigenstates  for the 
independent electron approximation}. Since the ground-state in the interacting
case, even for infinitesimal interaction, changes qualitatively in comparison 
to the non-interacting case, Eqs.(\ref{E19},\ref{E20}) cannot be connected in 
a perturbative way to the ground-state of the disordered but
non-interacting system. 

The GSs presented above are strongly degenerated. Their degeneracy at quarter
filling is given by the $N$ arbitrary $v_i$ values and the arbitrary (but 
non-zero) $w$, is proportional to the size of the system. The existence of 
such type of states for 2D Penrose type lattices has been first conjectured by
Semba and Ninomiya \cite{ra10} and Kohmoto and Sutherland \cite{ra7}, and
further analyzed in \cite{ra9,ra11}. From the here reported results it can be 
seen that this property is present also for other systems as well in the 
multielectronic and interacting case too, at least for $N=N_{\Lambda}$. We
stress however, that in the case $N < N_{\Lambda}$, the degree of the 
degeneracy strongly increases given as well by the geometrical degeneracy 
present in Eq.(\ref{E20}). The order of magnitude of the degeneration becomes
in this case $N_R=N_{\Lambda} ! /[N ! (N_{\Lambda} - N) !]$.

\section{Ground-state expectation values}

\subsection{The localized case.}

Despite the possibility to chose the Hamiltonian parameters in a 
spin-dependent way, the obtained GS is globally paramagnetic. At $1/4$ filling
($N=N_{\Lambda}$), the GS contains rigorously one electron on each site, so 
the hopping is completely forbidden in GS, and as a consequence, the system is 
localized, holding long-range density-density correlations. 

Indeed, 
calculating the ground-state expectation values through Eq.(\ref{E19}),
in this case we find for arbitrary $i \ne j$ and all $\sigma,\sigma'$
\begin{eqnarray}
\langle \hat d^{\dagger}_{i,\sigma} \hat d_{j,\sigma'} \rangle = 0, \quad
\langle \hat f^{\dagger}_{i,\sigma} \hat f_{j,\sigma'} \rangle = 0, \quad
\langle \hat d^{\dagger}_{i,\sigma} \hat f_{j,\sigma'} \rangle = 0, \quad
\langle \hat f^{\dagger}_{i,\sigma} \hat d_{j,\sigma'} \rangle = 0 ,
\label{E21}
\end{eqnarray}
where $\langle ...\rangle = \langle \Psi_g(N_{\Lambda})|...|\Psi_g(N_{\Lambda})
\rangle / \langle \Psi_g(N_{\Lambda})|\Psi_g(N_{\Lambda})\rangle$, 
$|\Psi_g(N_{\Lambda})\rangle$ is presented in Eq.(\ref{E19}), and
$\langle \Psi_g(N_{\Lambda})|\Psi_g(N_{\Lambda})\rangle = \prod_{i=1}^{N_{
\Lambda}} [(1+|w|^{-2})(1+|v_i|^2)]$. The reason for Eq.(\ref{E21}) is 
simple: $|\Psi_g(N_{\Lambda})\rangle$ contains exactly one electron on each 
lattice site, so $|\Psi_1(p,p')\rangle =  \hat p^{\dagger}_{i,\sigma} 
\hat p'_{j,\sigma'}|\Psi_g(N_{\Lambda})\rangle$, where $p,p'=d,f$ and 
$i\ne j$, contains a double occupancy, and as a consequence $|\Psi_1(p,p')
\rangle$ and $|\Psi_g(N_{\Lambda})\rangle$ are orthogonal. We underline that
since Eq.(\ref{E21}) holds for arbitrary $v_i$, it remain the same
after the average over the disorder ($v_i$ variables). Denoting the 
translational invariant averages by $\langle \langle ...\rangle \rangle =
\int P(\{ v_i \} ) (\prod_i d v_i)\langle...\rangle$, where $P(\{ v_i \})$ 
describes the distribution of the disordered variables
(being here arbitrary) and $\int P(\{ v_i \} ) 
(\prod_i d v_i) =1$ holds by definition, Eq.(\ref{E21}) automatically implies
as well $\langle \langle \hat p^{\dagger}_{i,\sigma} \hat p'_{j,\sigma'}
\rangle \rangle = 0$ for all $p,p'=d,f$, all $\sigma,\sigma'$ and all 
$i \ne j$. 

Furthermore, introducing for $i \ne j$ the notation
$D(i,j) = [(1+|w|^{-2})(1+|v_i|^2)(1+|v_j|^2)]^2$, we find
\begin{eqnarray}
&&\langle \hat n^d_{i,\sigma} \hat n^d_{j,\sigma} \rangle = D(i,j), \quad
\langle \hat n^d_{i,-\sigma} \hat n^d_{j,-\sigma} \rangle = |v_i|^2 |v_j|^2
D(i,j), 
\nonumber\\
&&\langle \hat n^f_{i,\sigma} \hat n^f_{j,\sigma} \rangle = |w|^{-4}D(i,j), 
\quad
\langle \hat n^f_{i,-\sigma} \hat n^f_{j,-\sigma} \rangle = \frac{|v_i|^2
|v_j|^2}{|w|^4} D(i,j), 
\nonumber\\
&&\langle \hat n^d_{i,\sigma} \hat n^f_{j,\sigma} \rangle = |w|^{-2} D(i,j), 
\quad
\langle \hat n^d_{i,\sigma} \hat n^f_{j,-\sigma} \rangle = \frac{|v_j|^2}{
|w|^2} D(i,j), 
\nonumber\\
&&\langle \hat n^d_{i,-\sigma} \hat n^f_{j,-\sigma} \rangle = \frac{|v_i|^2
|v_j|^2}{|w|^2} D(i,j), \quad
\langle \hat n^d_{i,-\sigma} \hat n^f_{j,\sigma} \rangle = \frac{|v_i|^2}{
|w|^2} D(i,j) .
\label{E22}
\end{eqnarray}
Starting from Eq.(\ref{E22}), for $\hat n_i = \sum_{\sigma} (\hat n^d_{i,
\sigma} + \hat n^f_{i,\sigma})$, based on Eq.(\ref{E22}) one obtains
\begin{eqnarray}
\langle \hat n_{i} \hat n_{j} \rangle = 1, \quad
\langle \langle \hat n_{i} \hat n_{j} \rangle \rangle = 1,
\label{E23}
\end{eqnarray}
where the second equality holds as explained below Eq.(\ref{E21}). 
Introducing now $\hat S_i^z =(1/2)[(\hat n^d_{i,\uparrow} + \hat n^f_{i,
\uparrow}) - (\hat n^d_{i,\downarrow} + \hat n^f_{i,\downarrow})]$, based 
again on Eq.(\ref{E22}), for $i \ne j$ we have
\begin{eqnarray}
\langle \hat S^z_i \hat S^z_j \rangle = \frac{p_i p_j}{4} , 
\label{E24}
\end{eqnarray} 
where $p_n = (1-|v_n|^2)/(1+|v_n|^2)$ takes arbitrary values in the domain
$(-1,+1)$, so $\langle \langle \hat S^z_i \hat S^z_j \rangle \rangle =0$
arises. As can be observed, $|\Psi_g (N_{\Lambda})\rangle$ indeed describes
a paramagnetic, completely localized ground-state containing long-range
density-density correlations.  At quarter filling, since $|\Psi_g (N=N_{
\Lambda})\rangle$ coherently controls the particle number occupancy at all 
lattice sites forbidding the hopping (and non-local hybridization) in the same
time, the GS clearly presents phase coherence over the whole lattice.

\subsection{The delocalized case.}

Under quarter filling, empty sites emerge in the GS, Eq.(\ref{E21}) deduced
through Eq.(\ref{E20}) does not hold, hopping is no more forbidden, and as a 
consequence, a delocalization occurs, the system becoming itinerant (remaining
further paramagnetic). Indeed, in this case, at $N < N_{\Lambda}$, based on
Eq.(\ref{E20}), the GS wave function can be written as
$|\Psi_g(N)\rangle = \sum_{R_{N}} \alpha_{R_N} |\Psi (R_N)\rangle
$, where $|\Psi (R_N)\rangle \equiv
|\Psi_{R_N}(\{ v_i \} )\rangle = \sqrt{D^{-1}_{R_N}(\{v_i\})}
\prod_{j\in R_N}({\hat {\bar O}}^{\dagger}_{j,\sigma} + v_j {\hat {\bar O}}^{
\dagger}_{j,-\sigma})|0\rangle$ build up an ortho-normalized wave function set 
containing $N_R$ components, we have $D_{R_N}(\{v_i\})=(1+|w|^{-2})^N \prod_{
j \in R_N} (1+|v_j|^2)$, and $\alpha_{R_N}$ are coefficients independent on 
the disordered $\{v_i\}$ set. The operators of the type $\hat p^{\dagger}_{i,
\sigma} \hat p'_{j,\sigma}$, where $p,p'=d,f$, now have non-zero matrix 
elements 
between ground-state components $|\Psi(R_N)\rangle, |\Psi(R'_N)\rangle$ 
describing $R_N,R'_N$ domains of the form $R_N=D_{N-1}+{\bf i}$, $R'_N =
D_{N-1}+{\bf j}$, where $D_{N-1}$ represents an arbitrary region of the 
lattice containing $N-1$ lattice sites, and ${\bf i},{\bf j}$ are representing
two different but arbitrary (not necessarily nearest-neighbor) sites of the
lattice. We have for example
\begin{eqnarray}
&&\langle \Psi(R_N)| \hat d^{\dagger}_{i,\sigma} \hat d_{j,\sigma}|\Psi (R'_N)
\rangle = \frac{(1+|w|^{-2})^{-1}}{\sqrt{(1+|v_i|^2)(1+|v_j|^2)}} \: , 
\nonumber\\
&&\langle \Psi(R_N)| \hat f^{\dagger}_{i,\sigma} \hat f_{j,\sigma}|\Psi (R'_N)
\rangle = \frac{(1+|w|^{2})^{-1}}{\sqrt{(1+|v_i|^2)(1+|v_j|^2)}} \: .   
\label{E25}
\end{eqnarray}
Since the disordered variables emerge in Eq.(\ref{E25}) through $|v_i|,|v_j|$
non-negative numbers, the average over the disorder maintains the non-zero
values in Eq.(\ref{E25}). As a consequence, the hopping being no more 
forbidden, the system becomes indeed itinerant. Since as seen from 
Eq.(\ref{E25}) all $d$ or $f$ electrons can hop everywhere in the ground-state,
the wave function in Eq.(\ref{E20}) is clearly an extended state. The 
conducting nature of the extended states can be demonstrated (see
for example [\cite{geb}]) through the variation of the chemical potential as 
the number of electrons vary. For this reason, let us observe that the 
ground-state wave function from Eq.(\ref{E20}) acting on the Hamiltonian from
Eq.(\ref{E7}), by the construction of the wave function as explained in 
Sect.V., has the property $\hat H |\Psi_g ( N < N_{\Lambda} ) \rangle =  
E_g(N)  |\Psi_g ( N < N_{\Lambda} ) \rangle = 0$, where $E_g(N)$ is the 
ground-state energy for $N$ particles in the system. Since $|\Psi_g ( N < N_{
\Lambda} ) \rangle$ is a wave function with non-zero norm, this relation means
$E_g(N)=0$. As a consequence, for $N \leq N_{\Lambda}-1$, we have for 
$\mu^{+}= E_g(N+1) - E_g(N)$ and $\mu^{-} = E_g(N) - E_g(N-1)$, the expression
\begin{eqnarray} 
\mu^{+} - \mu^{-} = 0 .
\label{eqmu}
\end{eqnarray}
Therfore, the state we analyze is conducting (see also [\cite{geb1}]).

Furthermore, the $\hat S^z_i \hat S^z_j$ operator will have non-zero matrix 
elements only along the diagonal in $R_N$ variables, and in conditions
mentioned for $p_i$ after Eq.(\ref{E24}) we further have $\langle \langle
\hat S^z_i \hat S^z_j \rangle \rangle = 0$. 

As can be seen, Eqs.(\ref{E19},\ref{E20}) describe a 
localization-delocalization transition driven by the electron concentration 
$\rho_n$, which emerge at $\rho^c_n=1/4$, the delocalized phase being present 
in the region $\rho_n < \rho^c_n$. The occurrence of this transition is 
intimately connected to the multielectronic nature of the description which is
made in the presence of the inter-particle interaction and absence of trapping
centers. Indeed, the problem considered at the level of independent electron 
approximation (e.g. absence of inter-electronic interaction) in presence of  
trapping centers leads to a one-particle problem in the presence of an 
attractive potential, which ends up usually at small energies with 
localization. Here all these are avoided.

Concerning the possibility of the emergence of Griffiths phases in influencing
the described transition, we mention that the Griffiths singularities arise
due to the presence of statistically rare clusters that are anomalously 
strongly coupled, and hence they are unique features of the disordered system
(see for example [\cite{geb2}]). The effect becomes weaker with increasing
dimension, increasing interaction, increasing the number of the components
$\bar N$ of the dynamical variables. In our case $\bar N=3$ (for example for 
spin), the dimension of the (quantum mechanical) description is $D=2$, and the 
results are not valid at zero inter-electronic interaction. All these 
conditions make unlikely the major influence of Griffiths phases, especially 
when the $p_i$ parameters are all maintained perfectly random at all 
sites as mentioned below Eq.(\ref{E24}), prohibiting in this way the local
formation of anomalously strongly coupled clusters.

\section{Summary and Conclusions.}

We deduced exact multielectronic concentration dependent ground-states
for disordered and interacting two-dimensional quantum mechanical systems at 
and below quarter filling. The ground-states describe a 
localization-delocalization transition driven by concentration and provide 
paramagnetic behavior. The ground-state nature is lost in the absence of the 
interaction e.g. independent electron approximation. The deduced results are 
non-perturbative and cannot be perturbatively reached from the 
non-interacting, altough disordered case. The studied system is of 
two band type, and the disorder is present independently in both 
$\hat H_{int}$ and $\hat H_0$ parts of the Hamiltonian, the trapping centers
being excluded. The presented procedure extends the exact solution 
possibilities known in 2D Penrose lattice case to non-quasicrystalline 
disordered systems as well, even in the presence of the inter-electronic 
interactions.

\section{Acknowledgements}

The research was supported by the Hungarian Fund for Scientific Research
through contracts OTKA-T-037212 and FKFP-0471. The author thanks to Florian
Gebhard for valuable discussions relating metal-insulator transitions.
 
\newpage

\begin{figure}[h]
\centerline{\epsfbox{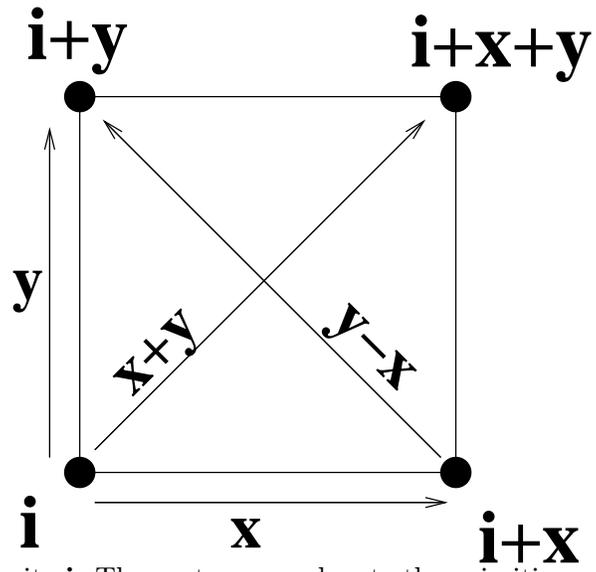}}
\caption{Unit cell $I$ at site ${\bf i}$. The vectors ${\bf x},{\bf y}$ denote
the primitive vectors of the unit cell, and arrows indicate the possible 
${\bf r}$ values allowed for the hoppings.}
\label{fig1}
\end{figure}

\newpage

\begin{figure}[h]
\centerline{\epsfbox{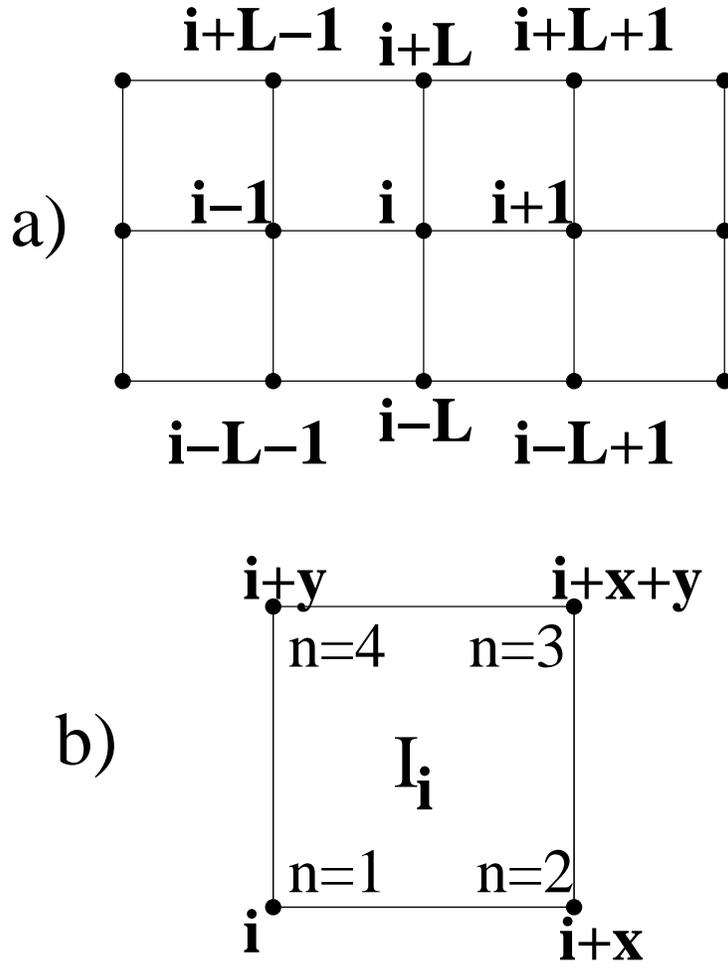}}
\caption{(a) The numbering of sites in a $L \times L$ two dimensional lattice
around the lattice site ${\bf i}$, and 
(b) The unit cell $I_{\bf i}$ placed at an arbitrary lattice site 
${\bf i}$, together with the ${\bf i}$ independent notation $(n=1,2,3,4)$ of
sites inside $I_{\bf i}$. $({\bf x},{\bf y})$ are the primitive vectors of the
unit cell.}  
\label{fig2}
\end{figure}

\newpage

\begin{figure}[h]
\centerline{\epsfbox{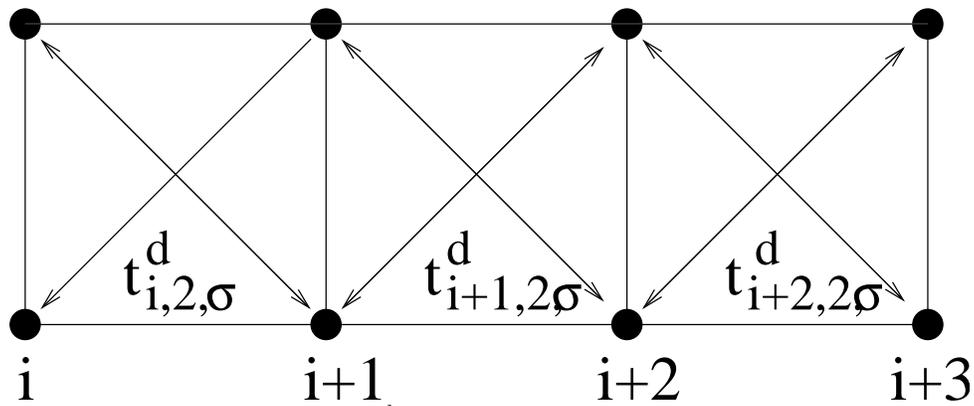}}
\caption{The independent diagonal $t^d_{i,2,\sigma}$ hopping amplitudes 
indicated by arrows in different unit cells.}
\label{fig3}
\end{figure}

\newpage

\begin{figure}[h]
\centerline{\epsfbox{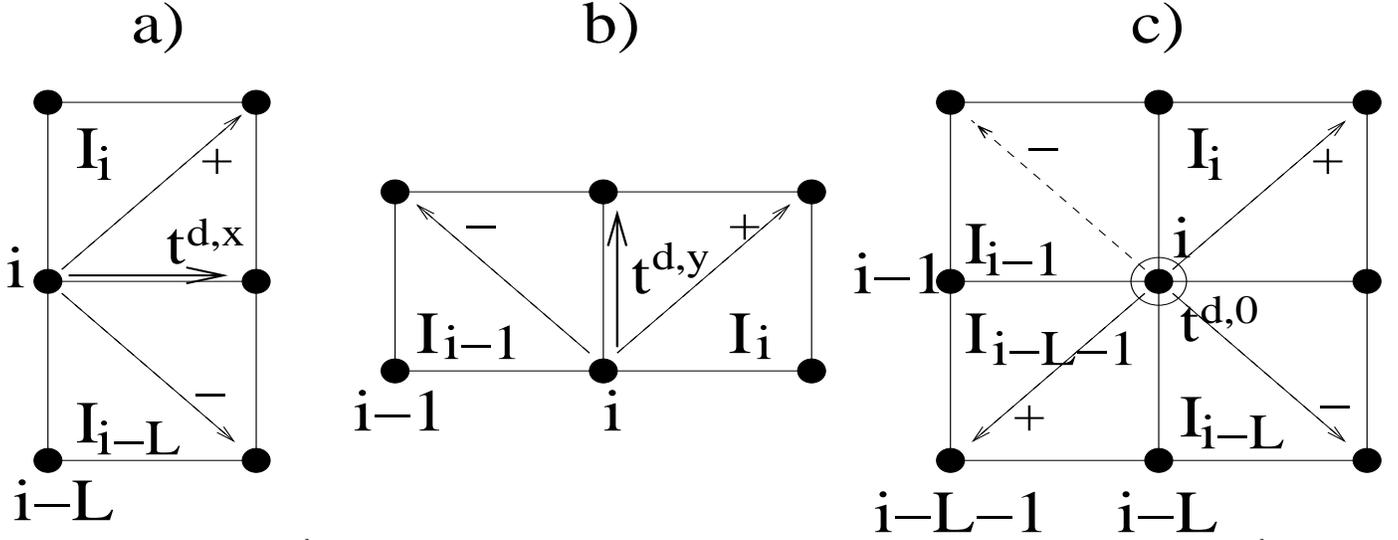}}
\caption{The $t^{d,\nu}_{i,\sigma}$ amplitudes at site $i$ for a) $\nu=x$, 
b) $\nu=y$, and c) $\nu=0$ respectively. $t^{d,\nu}_{i,\sigma}$ is denoted by 
full arrow in a),b), and by a circle in c). Dotted arrows with $\pm$ label
indicate the $t^{d,\pm}_{j,\sigma}$ random amplitudes that enter in the 
expression of $t^{d,\nu}_{i,\sigma}$ presented in the plots a),b),c) from 
Eq.(\ref{E13}). In all plots the notation of the unit cell $I_j$ at site $j$,
containing $t^{d,\nu}_{j,\sigma}$ is also presented. For example, in a) $I_i$
defined at site $i$ contains $t^{d,+}_{i,\sigma}$, and $I_{i-L}$ defined at the
site $i-L$ contains $t^{d,-}_{i-L,\sigma}$.}
\label{fig4}
\end{figure}

\end{document}